# 6. Assessing employment and labour issues implicated by using AI

*Thijs Willems, Darion Jin Hotan, Jiawen Cheryl Tang, Norakmal Hakim bin Norhashim, King Wang Poon, Zi An Galvyn Goh, Radha Vinod*

**This manuscript is accepted for publication in *Emad Yaghmaei, et al., eds., Global Perspectives on AI Impact Assessment (Oxford University Press, forthcoming 2025).***

## 1. Introduction

Emerging technologies operating on Artificial Intelligence (AI) and their effects on the employment and labour of human workers has received a tremendous amount of attention in recent years. However, most existing studies approach this issue from a 'reductionist view on work'[1], : tasks and skills are assumed to be bounded and independent elements of an occupation that can be replaced or augmented by AI rather unproblematically. This chapter challenges this assumption, arguing that tasks and skills are highly interdependent aspects of work, so that a more systemic view with more granular data on work is warranted to assess AI's impact. We offer two approaches that put forward this systemic understanding of work to advance academic debates and practical decisions around AI assessment: 1) the importance of context in work, and 2) a reconsideration of tasks as interdependent units of analysis.

With the ongoing Fourth Industrial Revolution, digitalisation and its scale of influence on the economy, workplaces, and employment and labour issues are well acknowledged by now[2], [3], [4]. Discussions about the increasing use, application and efficiency of technologies and the speed of digital disruption in all spheres of life have been the centre of attention. Many studies converge around advances in Artificial Intelligence (AI), such as machine learning[5], [6], algorithmic technologies[7], [8] and the potential impact of generative AI and large-language models[9]. Although the predicted consequences of disruption by AI are diverging and range from the threat of impending mass-unemployment[10] to more moderate and nuanced observations that automation and new technologies may create whole new industries[11], [12] the large majority of studies and reports agree on the necessity to rethink work and labour in profound new ways.

---

[1] Bailey, D. E., & Barley, S. R. (2020). Beyond design and use: How scholars should study intelligent technologies. *Information and Organization*, *30*(2), 100286.
[2] Bailey, D., Faraj, S., Hinds, P., von Krogh, G., & Leonardi, P. (2019). Special issue of organization science: Emerging technologies and organizing. Organization Science, 30(3), 642-646.;
[3] Raisch, S., & Krakowski, S. (2021). Artificial intelligence and management: The automation–augmentation paradox. Academy of Management Review, 46(1), 192-210.
[4] Schwab, K. (2017). The fourth industrial revolution. Crown Currency; Susskind, D. (2020). *A world without work: Technology, automation and how we should respond*. Penguin UK.
[5] Brynjolfsson, E., Mitchell, T. and Rock, D., (2018), May. What can machines learn, and what does it mean
for occupations and the economy?. In *AEA Papers and Proceedings* (Vol. 108, pp. 43-47).
[6] Faraj, S., Pachidi, S., & Sayegh, K. (2018). Working and organizing in the age of the learning algorithm. *Information and Organization*, *28*(1), 62-70
[7] Kellogg, K. C., Valentine, M. A., & Christin, A. (2020). Algorithms at work: The new contested terrain of control. *Academy of Management Annals*, *14*(1), 366-410.;
[8] Willems, T., & Hafermalz, E. (2021). Distributed seeing: Algorithms and the reconfiguration of the workplace, a case of 'automated' trading. *Information and Organization*, *31*(4), 100376.
[9] Brynjolfsson, E., Li, D., & Raymond, L. R. (2023). *Generative AI at work* (No. w31161). National Bureau of Economic Research.

An impending disruption of the future of work brings uncertainty and sometimes fear. In response, a powerful discourse has emerged that frames the future of work in a specific technological vein. Technology is framed as an unstoppable force we have little control over[13],[14] and our adaptability in terms of up- and re-skilling is the primary solution to thrive in this new reality[15]. Such framings are not neutral representations of reality, but they are performative in the sense that they participate in shaping the future that they describe. In the case of AI's impact on work and labour, the future is imprinted as one where technology has power and human agency is rendered to the level of responding and adapting via national upskilling programs. Yet, this is a one-dimensional framing of the future and, as a result, does not benefit a holistic assessment of the impact of AI on work and labour. While up- and re-skilling programs can hardly be seen as a negative response to such impact assessments, the problem is that it is framed as the only solution. This creates its own vulnerabilities as well as potential unintended consequences, specifically because the trickle down effect of overemphasising new skills is not considered.

This chapter takes stock of the discussion above by drawing closer to the realities of workplaces in which many of the transformations fuelled by AI actually happen. Previous literature as well as popular media tend to either emphasise automation, where human expertise is replaced by technology (e.g. because it is more efficient), or augmentation, where humans and machines focus on tasks they are individually better at[16]. Both of these perceptions however are limiting as they fail to capture the realities of the modern workplace. We argue that a technological deterministic frame that primarily focuses on the new - often digital - skills that are necessary for the workforce to thrive, reflects a common misperception about how technology and human workers interrelate. Barley and Beane term this the 'reductionist view of work', highlighting that research often tends to reduce complex jobs to a set of individual, unrelated tasks that can then be assessed and re-engineered rather instrumentally[17]. While a technology may automate a specific task, this does not automatically mean that work will become redundant as this remains a negotiation[18].

---

[10] Ford, M. (2016). The rise of the robots: Technology and the threat of mass unemployment. *International Journal of HRD Practice Policy and Research*, *111*

[11] Autor, D., (2015). Why are there still so many jobs? The history and future of workplace automation. Journal of Economic Perspectives, 29(3), pp. 3-30.;.

[12] Wajcman, J. (2017). Automation: is it really different this time?. The British Journal of Sociology, 68(1), pp. 119-127.

[13] Urry, J. (2016). What is the Future?. Cambridge: Polity.

[14] Waardenburg, L., Huysman, M., & Sergeeva, A. V. (2022). In the land of the blind, the one-eyed man is king: Knowledge brokerage in the age of learning algorithms. Organization Science, 33(1), 59-82

[15] Schlogl, L., Weiss, E., & Prainsack, B. (2021). Constructing the 'Future of Work': An analysis of the policy discourse. *New Technology, Work and Employment*, *36*(3), 307-326.

[16] Raisch, S., & Krakowski, S. (2021). Artificial intelligence and management: The automation–augmentation paradox. Academy of Management Review, 46(1), 192-210.

[17] Barley, S. R., & Beane, M. (2020) 'Intelligent Technologies' Implications for Work', in Barley, S. R. (ed.) *Work and Technological Change.* Oxford: Oxford University Press, pp. 69–115.

[18] Barley, S. R., & Beane, M. (2020) 'Intelligent Technologies' Implications for Work', in Barley, S. R. (ed.) *Work and Technological Change.* Oxford: Oxford University Press, pp. 69–115.

In drawing closer to the realities of how AI technologies are implemented in organisations and impact work and labour, this chapter proposes two different yet complementary approaches that build on recent literature in information technology and workplace studies as well as prior research conducted by the authors. Our main goal is to address the critique offered by Barley and Beane (2020) on a reductionist view of work[19] and to, instead, study AI firmly within the context of the workplace. Rather than reducing work to a set of tasks or roles, we aim to show how tasks and roles form an intricate system with other tasks, roles, responsibilities, workplace situations, professions, and are even interrelated to entire industries and value systems. This contributes to AI impact assessments by offering a more holistic understanding of how AI transforms work and labour. Specifically, we argue, it allows for assessing the impact of AI on the level of human expertise, and how this changes the roles and interactions professionals experience at work. This can inform organisational and national policies beyond addressing dominant issues in the current discourse on the future of work to, instead, provide a more human-centric assessment on the potential of AI elevating rather than reducing the role of human labour.

## 2. Approach 1: Back to the thick of it

A key issue that Barley and Beane discuss within current reductionist views of work prevalent in a majority of studies on intelligent technologies is that a direct, causal relationship is assumed between employment and the automation of a skill, task, or job[20]. This perspective regards jobs as independent entities that are disassociated from each other, or as neatly bounded activities that can be replaced or complemented by technology. From this assumption, a single task within a job can be automated or replaced without affecting other associated tasks. This is problematic, they propose, because it: first, portrays the effects of AI as monolithic without giving due attention to context; second, because existing empirical work does not suggest many overarching consistencies in how technology occasions change in employment; and third, because this approach is less sensitive to uncovering and understanding differences and similarities across contexts of AI use[21]. These are important issues for AI impact assessment as they lead to, at best, generalised recommendations and frameworks glossing over professional, organisational or industrial details shaping how AI is implemented and used.

In reality, individual jobs and tasks exist within a broader social network and are connected to each other through different relations. As such, tasks are embedded as a part of their own work practices and work contexts. At the same time, these work practices are attached to roles and these roles are also embedded in and constitutive of role relations that define networks of roles. A more contextual understanding of AI implementation and use is thus warranted in order to effectively map out AI impact assessments[22]. Recent research has for example suggested that rather than looking at how AI impacts work and technology - either in beneficial or in dark ways - a contextualised understanding of work problematizes this unidirectional relationship because human workers and intelligent technologies actually seem to become more entangled and interdependent in practice[23], [24].

---

[19] Barley, S. R., & Beane, M. (2020) 'Intelligent Technologies' Implications for Work', in Barley, S. R. (ed.) *Work and Technological Change.* Oxford: Oxford University Press, pp. 69–115.
[20] Barley, S. R., & Beane, M. (2020) 'Intelligent Technologies' Implications for Work', in Barley, S. R. (ed.) *Work and Technological Change.* Oxford: Oxford University Press, pp. 69–115.
[21] Barley, S. R., & Beane, M. (2020) 'Intelligent Technologies' Implications for Work', in Barley, S. R. (ed.) *Work and Technological Change.* Oxford: Oxford University Press, pp. 69–115.
[22] Barley, S. R., & Beane, M. (2020) 'Intelligent Technologies' Implications for Work', in Barley, S. R. (ed.) *Work and Technological Change.* Oxford: Oxford University Press, pp. 69–115.

A reductionist view of work would overlook these different but interdependent realms which could result in a superficial and sweeping overview in assessing technology's effects on work and employment. This approach has also resulted in different authors arriving at strikingly different conclusions about AI and its impact on work and employment, ranging from rather optimistic views on job displacement to dystopian predictions. In contrast, an approach that takes the context of AI implementation in the workplace seriously allows focusing on what people do when they use technology, with whom they interact, and how those interactions unfold and contribute to a particular form of expertise that is relational and has the potential to elevate the role of human workers[25], [26]. This makes researchers record how and why technologies give rise (or not) to organisational change from the ground up rather than an outside-in approach. Therefore, a non-reductionist view of work would leave ample room for different possibilities to occur, acknowledging even the possibility that the same technology may affect organisations in different ways whilst being open to the idea that there may be common outcomes and general patterns of change. Below, we draw on recent literature to illustrate what such thick descriptions of the impact of AI on work and labour looks like and how this contributes to more fine-grained impact assessments.

## 2.1 The importance of context when assessing AI impact

To explain what we described above, we now turn to some recent studies on the effects of AI technologies on labour and work that have explicitly aimed for countering the tendency of a reductionist approach to work. These studies, specifically, are methodologically sensitive to the context in which work and AI implementation occurs, attempting to study how human workers and technology interact and how this process shapes how technology is used and the effects it creates. Noteworthy, these are often ethnographic workplace studies operating from an interpretive research paradigm in order to explain AI use through detailed qualitative case descriptions. We highlight two main insights we observe in this body of work that are most relevant to this chapter.

First, workplace ethnographies have challenged the superiority of AI technologies from a knowledge perspective[27]. Such studies claim that the expertise of human workers and that of AI technologies should be understood from a systems perspective, i.e. with a focus on human-AI collaboration more so than automation or augmentation. Even with recent advances such as in generative AI this argument remains, with a recent review stating that human-AI collaboration remains a core theme explored in the literature and that future research should give more attention to how effective human-AI collaboration systems can be designed[28]. These studies address the assumption that automation of a plethora of tasks through AI would inherently be better because they run on neutral, objective, and massive amounts of data. Mayer-Schönberger and Cukier, for example, state that such automated decisions are superior to those humans can make because AI has access to and is capable of analysing such immense bodies of data so that correlations can be drawn with a predictive power that "will routinely be used to disprove our causal intuitions"[29]. While this observation is not invalid per se, it fails to capture the point that there is more to the work of professionals than simple cause and effect. The insight that certain tasks can be automated will thus result in a change in labour and employment is typical of reductionist studies of work as they do not take the context of work into sufficient account. This provides fuel for predictions about the demise of professions and expertise[30]. Yet, these predictions are built on a narrow understanding of what a professional does and knows, and that individual tasks can be seen as separate from a more holistic practice.

---

[23] Glaser, V. L., Pollock, N., & D'Adderio, L. (2021). The biography of an algorithm: Performing algorithmic technologies in organizations. *Organization Theory*, *2*(2), 26317877211004609.;

[24] Grønsund, T., & Aanestad, M. (2020). Augmenting the algorithm: Emerging human-in-the-loop work configurations. *The Journal of Strategic Information Systems*, *29*(2), 101614.

[25] Poon, K.W., Willems, T., Liu, W.S.Y. (2023). The Future of Expertise: From Stepwise Domain Upskilling to Multifaceted Mastery. In: Lee, W.O., Brown, P., Goodwin, A.L., Green, A. (eds) International Handbook on Education Development in Asia-Pacific. Springer, Singapore. https://doi.org/10.1007/978-981-16-2327-1_42-1;

[26] Pakarinen, P., & Huising, R. (2023). Relational Expertise: What Machines Can't Know. *Journal of Management Studies*.

[27] Anthony, C., Bechky, B. A., & Fayard, A. L. (2023). "Collaborating" with AI: Taking a system view to explore the future of work. *Organization Science;* Pakarinen, P., & Huising, R. (2023). Relational Expertise: What Machines Can't Know. *Journal of Management Studies*.

As a response, Hadjimichael and Tsouka, for example, have addressed this assumption in their literature review on tacit knowledge[31]. In drawing on, amongst others, the work of Hubert and Stuart Dreyfus, they argue that knowledge is more than a set of rule-following behaviours and decisions[32], [33]. In strictly rational, bounded systems AI may perform well but in reality work is almost always more than that and is a matter of rules as well as contingent, tacit judgments. Hadjimichael and Tsoukas thus argue that while recent advances in AI may appear to make tacit knowledge less relevant as tasks can be automated, the opposite is true as professionals actually need greater contextual knowledge of their work in assessing and judging AI predictions[34]. They take the example of a medical professional, saying that while "AI may be able to diagnose skin cancer (...) the diagnosis does not mean anything to the AI system"[35]. Instead, a skin cancer diagnosis is more than the observation of a fact, but requires a professional's situational understanding, the ability to discern which facts are relevant and how to interpret them, and engage with a diagnosis and patients on a more experiential level.

Pachidi, Berends, Faraj and Huysman discussed this in the context of the introduction of algorithmic technology in a sales organisation[36]. They describe this process via a 'regime-of-knowing' lens suggesting that the implementation of AI is more of an epistemological struggle between professionals than a functional, top-down process. Specifically, they looked at meetings and discussions between technologists responsible for the AI and sales managers whose work was to be affected by it, observing that the implementation process was guided by struggles around "what is worth knowing, what actions matter to acquire this knowledge, and who has the authority to make decisions around those issues"[37]. While the AI was ultimately implemented this was not so much because the technology was necessarily better but because of a tactical mistake made by sales managers leading to them losing status in the organisation. In sum, workplace studies on AI aimed at challenging the superiority of technology from a knowledge perspective have done so by shedding light on the complex interrelations human workers and technologies form and how this shapes the process of AI implementation and its effects on labour and employment [38], [39].

---

[28] Bankins, S., et al. (2023). A multilevel review of artificial intelligence in organizations: Implications for organizational behavior research and practice. *Journal of Organizational Behavior*.

[29] Mayer-Schönberger, V., & Cukier, K. (2013). *Big data: A revolution that will transform how we live, work, and think*. Houghton Mifflin Harcourt, p.64.

[30] Susskind, R. E., & Susskind, D. (2015). *The future of the professions: How technology will transform the work of human experts*. Oxford University Press, USA.

[31] Hadjimichael, D., & Tsoukas, H. (2019). Toward a better understanding of tacit knowledge in organizations: Taking stock and moving forward. *Academy of Management Annals*, *13*(2), 672-703.

[32] Dreyfus, H., Dreyfus, S. E. 2000). *Mind over machine*. Simon and Schuster.; Reference Hubert and Dreyfus 2005 missing

[33] Dreyfus, H. L., & Dreyfus, S. E., (2005). Peripheral vision: Expertise in real world contexts. Organization studies, 26(5), 779-792.

[34] Hadjimichael, D., & Tsoukas, H. (2019). Toward a better understanding of tacit knowledge in organizations: Taking stock and moving forward. *Academy of Management Annals*, *13*(2), 672-703.

[35] Hadjimichael, D., & Tsoukas, H. (2019). Toward a better understanding of tacit knowledge in organizations: Taking stock and moving forward. *Academy of Management Annals*, *13*(2), 672-703.

[36] Pachidi, S., Berends, H., Faraj, S., & Huysman, M. (2021). Make way for the algorithms: Symbolic actions and change in a regime of knowing. *Organization Science*, *32*(1), 18-41.

A second insight we derive from these studies is that AI implementation may automate a specific task or solve a specific solution, but that in doing so new uncertainties are simultaneously created. In other words, in introducing AI to specific tasks, the workplace and what constitutes work is being reconfigured [40]. This has important implications for how we assess labour and employment issues because it indicates that continuous up- and re-skilling may not be the only solution that is required. For instance, roles, responsibilities and tasks may become redistributed with AI technology. Waardenburg, Huysman and Sergeeva studied the emergence of a new occupational group when the Dutch police introduced an algorithm facilitating the prediction of crimes [41]. To translate these opaque predictions, the role of a knowledge broker was created aiming to interpret and communicate data with police management, which ultimately gave these brokers power by guiding how police should perform in the field [42].

This is due to a number of qualities that constitute emerging technologies in comparison to older, existing technology. Advances in AI, robotics and sensors, and data analytics offer the potential to not just impact the human body but also the human mind and senses. This makes automation and even augmentation a less likely outcome. For instance, in healthcare the introduction of robotics has vastly changed how surgeons work and what is required of them in terms of skills and knowledge. Here, rather than replacing the human body, we see a much closer relationship between the human and the machine and one that is based on the blending of sensing capabilities [43]. This impacts far beyond just the practices of the workers and redefines issues such as coordination, occupational roles and responsibilities, and providing new interdependencies [44].

[37] Pachidi, S., Berends, H., Faraj, S., & Huysman, M. (2021). Make way for the algorithms: Symbolic actions and change in a regime of knowing. *Organization Science*, *32*(1), p. 20.
[38] Christin, A., (2020). The ethnographer and the algorithm: beyond the black box. Theory and Society, 49(5), pp.897-918. (Christin, 2020; Orlikowski, 2000) reference missing
[39] Orlikowski, W.J., (2000). Using technology and constituting structures: A practice lens for studying technology in organizations. Organization science, 11(4), pp.404-428.
[40] Willems, T., & Hafermalz, E. (2021). Distributed seeing: Algorithms and the reconfiguration of the workplace, a case of automated' trading. *Information and Organization*, *31*(4), 100376.
[41] Waardenburg, L., Huysman, M., & Sergeeva, A. V. (2022). In the land of the blind, the one-eyed man is king: Knowledge brokerage in the age of learning algorithms. *Organization Science*, *33*(1), 59-82.
[42] Waardenburg, L., Huysman, M., & Sergeeva, A. V. (2022). In the land of the blind, the one-eyed man is king: Knowledge brokerage in the age of learning algorithms. *Organization Science*, *33*(1), 59-82.
[43] Sergeeva, A. V., Faraj, S., & Huysman, M. (2020). Losing touch: an embodiment perspective on coordination in robotic surgery. *Organization Science*, *31*(5), 1248-1271.
[44] Sergeeva, A. V., Faraj, S., & Huysman, M. (2020). Losing touch: an embodiment perspective on coordination in robotic surgery. *Organization Science*, *31*(5), 1248-1271.

Hence, there is a need to examine how technology impinges on human expertise and vice versa, as well as how expertise is distributed across a network of different humans and technology in various ways. The introduction of technology will not simply lead to replacement or substitution of human tasks. It will also create new workflows and new roles for humans, requiring different and new types of human expertise to be developed. In sum, the idea that AI reconfigures work and the role of human expertise is insightful because the implementation of AI not only solves a specific problem but at the same time may create new challenges. Assessing the impact on labour and employment implicated by using AI needs to take this into account in providing effective organisational or national policies.

## 3. Approach 2: Towards a Relational Understanding of Tasks

The foregoing ethnographic methods provide thick data about how people work. This potentially has two trade-offs: first, highly contextual ethnographic data often focus on a single research site and can thus limit the scope for comparisons across different organisations, industries and technologies; second, ethnographic research tends to be resource intensive in terms of time and manpower. Thus, these methodologies are best complemented with approaches that enable analysis across jobs/sectors/technologies in a resource-efficient manner. Approach 1 is about 'zooming-in' - yielding thick data which is rich, particular, and contextual. Approach 2 is about 'zooming-out' - yielding a complementary description that is broader, general, and comprehensive.

In service of this aim, this section explores the use of tasks as a research focus. Studies concerning the impact of AI technologies on labour and employment issues have sometimes invoked a 'task approach' – using job tasks as units of analysis to evaluate technological impact and change. In the preceding section, we introduced Barley's reductionist critique of the task approach in broad strokes. At the heart of his critique is the question of whether the task approach can be made to accommodate the relational aspects of work; we believe the answer is yes. We begin by defining the task approach, and assessing its contribution to the study of AI and work. We then show how the task approach can evolve in light of Barley's reductionist critique.

### 3.1 Defining the task approach

A task is "a unit of work activity that produces output" [45]. The term 'task approach' may refer to a swathe of analytical frameworks which give prominence to tasks in various ways. Acemoglu and Autor [46] point to insufficiencies in the canonical production function in economics, preferring a production function in which job tasks figure as the fundamental units of production [47]. Tasks are employed here in the study of economic phenomena, including job polarisation [48] and international offshoring [49]. The task approach is sufficiently flexible to admit a wide range of applications: task data may be used to explore social issues, as with the allocation of tasks to immigrant communities [50], the management of patients' tasks in their daily activities during rehabilitation [51], and more recently, to studying remote work during the Covid-19 pandemic [52].

---

[45] Acemoglu, D. and Autor, D., (2011). Skills, tasks and technologies: Implications for employment and earnings. In *Handbook of labor economics* (Vol. 4, pp. 1043-1171). Elsevier., p. 1045.
[46] Acemoglu, D. and Autor, D., (2011). Skills, tasks and technologies: Implications for employment and earnings. In *Handbook of labor economics* (Vol. 4, pp. 1043-1171). Elsevier., p. 1045.
[47] Autor, D., (2013). The "task approach" to labor markets: an overview. *Journal for Labour Market Research*, *46*(3), pp.185-199.

## 3.2 Assessing the contribution of the task approach

At the broadest level, tasks are *what people do at work* - they are actions undertaken in work settings to achieve work-related outcomes. By thinking about occupations as bundles of tasks, we arrive at an important insight: AI technologies disrupt the workforce task-by-task, rather than job-by-job. This principle should hold true for as long as AI technologies are limited to narrow (rather than general) AI; they are more suitable for certain kinds of tasks than others.

Understanding *which* tasks are suitable for AI automation involves abstracting general features from particular tasks and building a taxonomy of task types. This approach was employed by Autor, Levy and Murnane to study the relationship between computerization and job skill demands, an approach motivated by the view that the distinctions between routine/non-routine and cognitive/manual tasks are relevant in appraising susceptibility to automation [53]. Similarly, Brynjolfsson, Mitchell and Rock built a set of measures to characterise tasks by their 'suitability for machine learning', from which they make inferences about the effects of machine learning on different segments of the workforce [54]. These measures will have to evolve to keep pace with new generations of AI technologies; as Susskind observes, new AI often does not emulate human problem-solving strategies, which forces us to update our beliefs about their limitations [55]. Nonetheless, tasks continue to be used in the study of emerging technologies; to take a recent example, Eloundou et al. study Large Language Models and their effects on the labour market, using work tasks as their base unit of analysis [56].

Thus, the task approach broadens AI impact assessment in three ways:

---

[48] Goos, M. and Manning, A., (2007). Lousy and lovely jobs: The rising polarization of work in Britain. *The review of economics and statistics*, *89*(1), pp.118-133.
[49] Blinder, A.S., (2009). How many US jobs might be offshorable?. *World Economics*, *10*(2), p.41. van den Broek, E., Sergeeva, A., & Huysman, M. (2021). WHEN THE MACHINE MEETS THE EXPERT: AN ETHNOGRAPHY OF DEVELOPING AI FOR HIRING. *MIS Quarterly*, *45*(3).
[50] Peri, G. and Sparber, C., (2009). Task specialization, immigration, and wages. *American Economic Journal: Applied Economics*, *1*(3), pp.135-69.
[51] Nott, M.T., Chapparo, C. and Heard, R., (2009). Reliability of the perceive, recall, plan and perform
system of task analysis: A criterion-referenced assessment. Australian Occupational Therapy Journal, 56(5), pp.307-314.
[52] Dingel, J.I. and Neiman, B., (2020). How many jobs can be done at home?. *Journal of Public Economics*, *189*, p.104235.
[53] Autor, D., Levy, F. and Murnane, R.J., (2003). The skill content of recent technological change: An empirical exploration. *The Quarterly journal of economics*, *118*(4), pp.1279-1333.
[54] Brynjolfsson, E., Mitchell, T. and Rock, D., (2018), May. What can machines learn, and what does it mean for occupations and the economy?. In *AEA Papers and Proceedings* (Vol. 108, pp. 43-47).
[55] Susskind, D. (2020). *A world without work: Technology, automation and how we should respond*. Penguin UK.
[56] Eloundou, T., Manning, S., Mishkin, P., & Rock, D. (2023). Gpts are gpts: An early look at the labor market impact potential of large language models. *arXiv preprint arXiv:2303.10130*.

1. *Bridging ethnography and economics* - the task approach allows detailed ethnographic data (e.g., on the types of tasks people do at work) to be linked with measurable economic variables (e.g., wage polarisation - see the study by Autor, Levy and Murnane [57]). This paves the way for testable theories on the effects of AI automation on the workforce/economy.
2. *Analysis at scale* - since AI automates work at the level of individual tasks (or task-types), emerging technologies can affect many sectors at once (for example - many administrative tasks such as documenting/recording/retrieving information have been partially or wholly automated; these tasks are present in every sector). Thus the task approach yields a comprehensive picture of the *scale* of AI automation across the workforce.
3. *Tasks as shared language* - Work experiences are diverse. Depending on who is describing them, they can be written with differing terms and turns of phrases. Having common vocabularies to describe these elements would ease discussions surrounding work. By describing work in terms of tasks, a shared language would enable work analysts to communicate their analysis with greater clarity. Moreover, a shared language is also necessary to define group-based descriptions that encode shared narratives.

## 3.3 Evolving the task approach

In his reductionist critique on work (introduced in the preceding sections), Barley raises some interesting questions about the limitations of the task approach. His critique draws on a distinction between the *relational* and *non-relational* aspects of work. According to this distinction, tasks and skills belong to the non-relational aspect of work, which encompasses the 'recurrent activities'[58] associated with an occupation; the relational aspects of work encompass the system of work practices, situations, roles, and role relations. Thus, when the task approach is applied to understanding the effects of technological change on work, we might expect its contributions to be limited to the non-relational domain. And while "technologically induced changes… may spill over into the role's relational elements" [59], this effect is at best secondary and contingent.

Thus, the key question is whether the task approach can be extended to encompass the relational aspects of work. We find in Barley's work an instructive idea for this type of extension: placing tasks within a *work process* framework [60].

The *work process* approach characterises tasks as steps in a process. Instead of viewing tasks as isolated units, tasks may be 'chained', i.e. one task's output might serve as input to another. A single worker may execute one or more task-chains, and the output of the entire chain may serve as input to another worker's chain(s). In this way, each task is understood to be a component of a larger work process. Although component tasks in isolation are non-relational, the larger work processes are relational: tasks are connected to other tasks, chains are connected to other chains. In sum, impact assessments of AI on individual, isolated tasks may be overly simplistic when they do not take the larger context of work into account. Thick descriptions of workplaces and work practices may thus contribute to task analysis by providing this context and illustrating how individual tasks, practices and roles usually form a network of tasks, practices and roles instead.

---

[57] Autor, D., Levy, F. and Murnane, R.J., (2003). The skill content of recent technological change: An empirical exploration. *The Quarterly journal of economics*, *118*(4), pp.1279-1333.

[58] Barley (2020) 'How Do Technologies Change Organisation', in Barley S. R. (ed.) *Work and Technological Change.* Oxford: Oxford University Press, p. 29.

[59] Barley (2020) 'How Do Technologies Change Organisation', in Barley S. R. (ed.) *Work and Technological Change.* Oxford: Oxford University Press, p. 30.

[60] Barley, S. R., & Beane, M. (2020) 'Intelligent Technologies' Implications for Work', in Barley, S. R. (ed.) *Work and Technological Change.* Oxford: Oxford University Press, pp. 69–115.

A candidate methodology for this approach is the *work process network* which Barley attributes to Pentland [61]. Pentland's work process network is constructed in the following way: actions are documented in 'threads', which consist of sequentially ordered actions along with a number of attributes (people, places, times) [62]. Computer software (e.g., 'ThreadNet') generates network diagrams from the given threads. In the resulting network diagram, nodes represent actions and vertices represent 'handoffs' between actions [63]. The useful relational structure here obtained is the handoff between tasks within the same job role and between different job roles. By tracing the structure of handoffs in the network, we potentially gain insight into the relationships between workers, teams, and leaders within the organisational context [64].

This evolution of the task approach dovetails with Approach 1. Traditionally, the task approach abstracts tasks from their work settings, treating them as independent units of analysis. This new approach uses ethnographic methods to bridge the gap between tasks as independent units, and tasks as contextually-bound steps in a localised work process.

## 4. Discussion: Assessing Impact Using Approach 1 and 2

Here, we briefly highlight some of our preceding studies in which we have employed our proposed research approaches in different contexts and in different permutations. Specifically, we draw on these examples to show how they open up possibilities to assess the impact of AI on a number of issues that are generally not discussed. Most analyses of AI typically look at and aim to predict which jobs are or will be affected and the percentage of jobs that is or will be affected. In this section we show that impact assessment of AI can be done on many different and equally important levels. Specifically, we believe that the following questions can also be addressed in impact assessments: i) An ethical question: Should a specific task or job be disrupted? ii) A human-centric question: Should a specific task or job remain in human hands? iii) A question related to mental health: Does the implementation of AI improve or worsen the well-being of workers? iv) A question around workplace relationships: How does AI change the interactions between different professionals and the technologies they employ?; v) And finally a question around quality and performance: How does AI affect the expertise of workers and what are the discernable advantages or disadvantages of this impact?

---

[61] Pentland, B.T., 2016. ThreadNet: Tracing and visualizing associations between actions. *IFIP Working Group 8.2, Dublin, Ireland*.

[62] Pentland, B.T., 2016. ThreadNet: Tracing and visualizing associations between actions. *IFIP Working Group 8.2, Dublin, Ireland*.

[63] Barley, S. R., & Beane, M. (2020) 'Intelligent Technologies' Implications for Work', in Barley, S. R. (ed.) *Work and Technological Change.* Oxford: Oxford University Press, p. 110.

[64] Goh, Z.A.G., Poon, K.W. (2022). A Task-Based Approach to Lifelong Learning, Well-Being, and Resilience in the Workplace of the Future. In: Lee, W.O., Brown, P., Goodwin, A.L., Green, A. (eds) International Handbook on Education Development in Asia-Pacific. Springer, Singapore. https://doi.org/10.1007/978-981-16-2327-1_72-1

## 4.1 Empowering workers through task analysis

Technology - including AI technology - is disrupting jobs tasks-by-tasks and not jobs-by-jobs. Moreover, people in vastly different occupations may share similar tasks (e.g. data processing; creating and sending out invoices; etc.). Looking at which tasks are disrupted by a specific technological development helps to provide greater granularity when assessing the impact of that technology, e.g. in terms of the speed and scale (across occupations) of disruption on jobs. It also allows for coming up with well-informed transition options for workers' whose jobs might be disrupted. A task analysis can thus empower individuals by providing them with clarity on future career options as well as concrete pathways on how to get there.

This allows us to come up with recommendations that go beyond merely looking at what is technologically feasible. For instance, in our project 'Live With AI' [65] we mapped the tasks of a number of jobs together with interviewing workers in these jobs in order to ask four questions:

1. Can a task be disrupted by AI?
2. When can the task be disrupted?
3. Should it be disrupted?
4. How does the role change?

As can be seen, these questions go beyond assessing the impact of AI as a technology and, instead, introduces more value-laden assessments too. This gave us greater insight into *how* we should consider the impact of AI. For instance, based on this study we identified different scenarios of how AI impacts specific occupations. In some cases, tasks were deemed to be done better or more efficiently when handled by AI, and in other cases we saw tasks where AI could complement the human worker. However, we also found scenarios where - even if AI *could* automate a task - it was deemed better that they remain in human hands.

For an account director in a creative agency, for example, advances in generative AI might make the automation of developing client relationships, or build and communicate presentations to clients, a real possibility in the near future. However, these tasks require creativity, expertise, and human interaction, and these are generally grounded in a deep understanding of one's domain context and human culture, something AI technologies lack. Keeping these tasks in human hands may then make the difference between successful creative agencies and auto-generated services.

These insights have increased in relevance with recent advances in generative AI. In the aforementioned study with the creative agency's account director, we assessed that a quarter or more of the tasks of creating designs, concepts, and layouts could be automated by AI, in contrast to prevailing discourse that creative work could not be automated [66]. Advances in generative AI have since favoured a higher degree of automatability for these tasks. As advances in AI continue to chip away at the tasks humans perform, the task approach offers a salient strategy to ensure we form an accurate view of the future ground our decisions, instead of being blindsided by prevailing presumptions.

---

[65] Apsley, H. L., Bin Norhashim, N. H., Goh, Z. A. G., Vinod, R., Lim, W. K., Ferreyrol-Alesi, E., Robinet, P., Gridel, M., & Poon, K. W. (2019). AI and work: Equal to the task in Live with AI 2019 white paper: How to empower humans amid the rise of artificial intelligence in society (pp. 36–63). Live with AI.

In a similar vein, we looked at this from an angle of workers' well-being. A key issue workers in the modern workplace face is having excessive job demands placed on them. This may deplete their mental and physical resources while preventing them from replenishing these resources [67]. This can lead to adverse consequences for the well-being of workers, such as burnout and reduced performance which can impact workers [68]. Disruptions from AI and technology can exacerbate these effects, creating further imbalances between job demands placed on workers, and their ability to store sufficient resources and replenish resources. To address this, we investigated in our research what types of tasks workers find engaging and/or exhausting [69]. These tasks can be linked to concrete work situations so as to understand its impact on individual workers, teams, leaders and organisations. For example, based on data collected from our prior and ongoing fieldwork, we found that some workers in some occupations which handle data preferred to *not* automate one of their tasks that would potentially be easiest to automate: data cleaning. AI advances have made tremendous progress in this regard, but for some workers who handle data, the data cleaning tasks provide an important moment in their day-to-day work to take a step back from their otherwise demanding job. Hence, it allows them to replenish resources they need for more cognitively demanding tasks. Moreover, data cleaning is also perceived as a way to connect to some of their core domain skills in data science and data analysis and to understand with greater clarity what is going on in their dataset. Choosing not to automate certain tasks may help people do their work better and promote their well-being, although further research will be needed with larger sample sizes to unpack the contextual nuances of different situations, jobs, teams, organisations and industries.

---

[66] Brynjolfsson, E., Mitchell, T. and Rock, D., 2018, May. What can machines learn, and what does it mean for occupations and the economy?. In *AEA Papers and Proceedings* (Vol. 108, pp. 43-47).

[67] Demerouti, E., Bakker, A. B., Nachreiner, F., & Schaufeli, W. B. (2001). The job demands-resources model of burnout. Journal of Applied psychology, 86(3), 499.

[68] Bakker, A. B., Demerouti, E., & Verbeke, W. (2004). Using the job demands-resources model to predict burnout and performance. *Human Resource Management: Published in Cooperation with the School of Business Administration, The University of Michigan and in alliance with the Society of Human Resources Management*, *43*(1), 83-104.

[69] Goh, Z.A.G. et al., 2021. Development of a novel method to investigate well-being and job quality in job tasks. In Singapore Conference of Applied Psychology 2021. Singapore Conference of Applied Psychology.; Goh, Z.A.G., Poon, K.W. (2022). A Task-Based Approach to Lifelong Learning, Well-Being, and Resilience in the Workplace of the Future. In: Lee, W.O., Brown, P., Goodwin, A.L., Green, A. (eds) International Handbook on Education Development in Asia-Pacific. Springer, Singapore. https://doi.org/10.1007/978-981-16-2327-1_72-1

## 4.2 Understanding the impact of AI on expertise and workplace interactions

Extending some of the previous insights, here we provide some examples of AI impact assessment on the level of workplace changes, specifically by looking at the new and sometimes unintended demands this places on human workers.

In one project we studied how systems of algorithmic trading impact the work of traders [70], something initially assumed to be susceptible to automation since it primarily consists of working with statistical data. However, in the analysis of their case study the impact of the new AI technology introduced was neither captured in paradigms of automation nor that of augmentation. The analysis offers, instead, a reconfiguration perspective [71] that shows how digital transformation efforts through algorithms bring into being particular vulnerabilities because technology and human workers become increasingly dependent on each other. In other words, while work is at the cusp of being transformed in unprecedented ways, we do not know and cannot know the outcome of this transformation completely and neither is it likely that upskilling alone is the solution to all. Technologies impact existing skills and job roles but these in turn also impact how technology is being implemented and used. A view that takes technology as both the means and the end, is likely to cause an increased misalignment with the tacit qualities of human expertise that remain crucial to make systems function.

We extended insights from this case-study in our multi-site project Mastery in a Digital Age [72] examining what expertise means to technical professionals in the Chemical Engineering and Precision Engineering sectors in Singapore. Rather than starting with the technologies itself, what their function is and how they operate, we started with the social setting of the workplace that was faced with a demand to work with new technologies. Thus, through observations of workplace practices of technical professionals and deep, interpretive interviews with them, we wanted to provide a bottom-up view on how human workers and technologies interact and what implications this has for professional expertise. One main insight this approach helped us uncover was that participants, when probed beyond the immediate purpose of new AI technology, realised that with increasing digitisation there is a tendency to emphasise the "digital skills" at the cost of forgetting domain fundamentals.

Technical professionals' awareness of AI technologies and the acknowledgement of AI taking over some of the tasks previously done by them led them to looking at how they could use their skills and expertise to add further value to the output produced by technology. One of our participants shared that knowing how technology is able to do certain data analysis tasks for him made him realise that his ability to interpret and communicate the information generated by technology to stakeholders became even more important. This substantiates our point that the adoption of AI technology does not merely replace or make redundant human expertise and labour. In fact, the advancement of AI and other technology has brought about the synchronous transformation and development within human expertise and skills, forcing individuals to refine and gain deeper mastery of their domain expertise and skills.

---

[70] Willems, T., & Hafermalz, E. (2021). Distributed seeing: Algorithms and the reconfiguration of the workplace, a case of 'automated' trading. *Information and Organization*, *31*(4), 100376.

[71] Suchman, L. A. (2007). *Human-machine reconfigurations: Plans and situated actions*. Cambridge university press.

[72] Poon, K.W., Willems, T., Liu, W.S.Y. (2023). The Future of Expertise: From Stepwise Domain Upskilling to Multifaceted Mastery. In: Lee, W.O., Brown, P., Goodwin, A.L., Green, A. (eds) International Handbook on Education Development in Asia-Pacific. Springer, Singapore. https://doi.org/10.1007/978-981-16-2327-1_42-1

Remarkably, we found that while AI may automate or augment certain work practices in some cases, assessing it only in this respect remains superficial at best. In fact, for the participants in our study, the introduction of digital technologies reconfigured their workplace and expertise to the extent that they now need to interact and relate with technology and other people *more and more intensely,* and not less. Technology may be able to bring about tangible benefits and lead to greater efficiency in work processes. But if we only conclude from this that this requires learning to operate these new technologies or making use of the results and analyses that technology provides us, we run the risk of losing out of sight the basic domain fundamentals, skills and tacit understandings of one's job. These are crucial to acquiring mastery in the digital age and assuring that human workers can do their work even better with AI.

# 5. Conclusions

This chapter proposed two approaches to studying work and AI that respond to the majority of existing studies on AI technology that tends to reduce work to a set of unrelated and general tasks. Our approaches, on the contrary, emphasise granularity in terms of contextual detail and relationality in terms of treating tasks and skills as interdependent. We have argued that ethnographic approaches can yield thick data about *how* people work when AI is introduced, and that our approach to tasks yields big data on *what* people do at work while acknowledging the relational quality of tasks. Below, we briefly highlight the potential possibility of combining these two approaches to provide a systemic and holistic understanding of AI's impact on work, and the implications of this on how labour and employment issues should be assessed by practitioners and policymakers.

The value of our argument that a systemic analysis of AI's impact on work is needed is clear when looking at some of the examples given above. Conventional approaches would have considered routine tasks as belonging to the category of the dull, dirty, and dangerous that would immediately be assessed as a task to be automated. Our alternative analyses we provided demonstrate that doing so could have had negative consequences on the wellbeing of workers, their agency in crafting their own career, as well as on their expertise and interactions. An increase in automation efficiency might paradoxically lead to poorer employment performance and wellbeing because of a decline in engagement, energy levels, and overall well-being.

Moreover, our chapter addresses the critique we put forward about current dominant discourses on the future of work, where upskilling and reskilling are seen as the only viable ways forward [73]. Imagine a worker who is asked to reskill and upskill because the tasks and relations that give their work purpose and are central to their expertise are now eliminated. They would feel alienated, and an inevitable initial resistance to the new skill or technology could be developed, potentially hardening into a more protracted resistance.

---

[73] Schlogl, L., Weiss, E., & Prainsack, B. (2021). Constructing the 'Future of Work': An analysis of the policy discourse. *New Technology, Work and Employment*, *36*(3), 307-326.

Thus, while certain tasks perhaps can be automated, a non-technologically deterministic view would argue that a more fundamental question should be whether we want a task to be automated or not. This can be a matter of protecting the wellbeing of workers by maintaining those aspects of work that energise, but it can also be a matter of keeping the human in the loop in safety critical operations so that core skills do not deteriorate. Automating for the sake of automating, because it is technologically possible, may lead to inferior outcomes that could feed further into employees' fear and sense of uncertainty, already prevailing in many workplaces, feeding a vicious cycle where performance could spiral downwards. Employers should also take this into account when designing new work processes augmented with AI: tasks and skills do not exist in isolation, and what seems like a simple replacement of a task or streamlining of a process may cut to the core of a worker's understanding of their job.

When extended across substantial segments of the economy and society, this could have broader socio-economic ramifications. The ramifications could be even political, given how contentious employment quality, unemployment, and underemployment have become in political debates. It could also exacerbate the loss of agency and control that workers and citizens are already feeling about decisions on technologies often taken by a small elite. Debates on the future of work, and especially on the impact of AI on workplaces, should take these issues into account. Without a systemic and holistic approach to undertake AI and work, and without superior outcomes for technology in the workplace, it is hard for anyone to claim that AI can be for good.

On the other hand, with a more systemic appreciation, we can be more discerning about what AI can do, what is desirable for AI to do, and what should remain in human hands. We can also better assess the impact of AI on the relationships between workers, especially the crucial relationships that nurture creativity, professionalism, and mutual learning.  We can determine where AI and humans can work in a distributed and complementary fashion, and what combinations of all these could give us a superior outcome both for the workplace and for the workers.

# References

Apsley, H. L., Bin Norhashim, N. H., Goh, Z. A. G., Vinod, R., Lim, W. K., Ferreyrol-Alesi, E., Robinet, P., Gridel, M., & Poon, K. W. (2019). AI and work: Equal to the task in Live with AI 2019 white paper: How to empower humans amid the rise of artificial intelligence in society (pp. 36–63). Live with AI.

Acemoglu, D. and Autor, D., 2011. Skills, tasks and technologies: Implications for employment and earnings. In *Handbook of labor economics* (Vol. 4, pp. 1043-1171). Elsevier.

Anthony, C., Bechky, B. A., & Fayard, A. L. (2023). "Collaborating" with AI: Taking a system view to explore the future of work. *Organization Science*

Autor, D., 2013. The "task approach" to labor markets: an overview. *Journal for Labour Market Research*, *46*(3), pp.185-199.

Autor, D., 2015. Why are there still so many jobs? The history and future of workplace automation. *Journal of Economic Perspectives*, *29*(3), pp. 3-30.

Autor, D., Levy, F. and Murnane, R.J., 2003. The skill content of recent technological change: An empirical exploration. *The Quarterly journal of economics*, *118*(4), pp.1279-1333.

Bailey, D., Faraj, S., Hinds, P., von Krogh, G., & Leonardi, P. (2019). Special issue of organization science: Emerging technologies and organizing. Organization Science, 30(3), 642-646.

Bailey, D. E., & Barley, S. R. (2020). Beyond design and use: How scholars should study intelligent technologies. *Information and Organization*, *30*(2), 100286.

Bakker, A. B., Demerouti, E., & Verbeke, W. (2004). Using the job demands-resources model to predict burnout and performance. *Human Resource Management: Published in Cooperation with the School of Business Administration, The University of Michigan and in alliance with the Society of Human Resources Management*, *43*(1), 83-104.

Bankins, S., Ocampo, A. C., Marrone, M., Restubog, S. L. D., & Woo, S. E. (2023). A multilevel review of artificial intelligence in organizations: Implications for organizational behavior research and practice. *Journal of Organizational Behavior*.

Barley (2020) 'How Do Technologies Change Organisation', in Barley S. R. (ed.) *Work and Technological Change.* Oxford: Oxford University Press, pp. 25–68.

Barley, S. R., & Beane, M. (2020) 'Intelligent Technologies' Implications for Work', in Barley, S. R. (ed.) *Work and Technological Change.* Oxford: Oxford University Press, pp. 69–115

Blinder, A.S., 2009. How many US jobs might be offshorable?. *World Economics*, *10*(2), p.41.

van den Broek, E., Sergeeva, A., & Huysman, M. (2021). WHEN THE MACHINE MEETS THE EXPERT: AN ETHNOGRAPHY OF DEVELOPING AI FOR HIRING. *MIS Quarterly*, *45*(3).

Brynjolfsson, E., Mitchell, T. and Rock, D., 2018, May. What can machines learn, and what does it mean for occupations and the economy?. In *AEA Papers and Proceedings* (Vol. 108, pp. 43-47).

Brynjolfsson, E., Li, D., & Raymond, L. R. (2023). *Generative AI at work* (No. w31161). National Bureau of Economic Research.

Christin, A., 2020. The ethnographer and the algorithm: beyond the black box. *Theory and Society*, *49*(5), pp.897-918.

Demerouti, E., Bakker, A. B., Nachreiner, F., & Schaufeli, W. B. (2001). The job demands resources model of burnout. Journal of Applied psychology, 86(3), 499.

Dingel, J.I. and Neiman, B., 2020. How many jobs can be done at home?. *Journal of Public Economics*, *189*, p.104235.

Dreyfus, H., Dreyfus, S.E., 2000. *Mind over machine*. Simon and Schuster.

Dreyfus, H. L., & Dreyfus, S. E., 2005. Peripheral vision: Expertise in real world contexts. *Organization studies*, *26*(5), 779-792.

Eloundou, T., Manning, S., Mishkin, P., & Rock, D. (2023). Gpts are gpts: An early look at the labor market impact potential of large language models. *arXiv preprint arXiv:2303.10130*.

Faraj, S., Pachidi, S., & Sayegh, K. (2018). Working and organizing in the age of the learning algorithm. *Information and Organization*, *28*(1), 62-70.

Ford, M. (2015). The rise of the robots: Technology and the threat of mass unemployment. *International Journal of HRD Practice Policy and Research*, *111*

Glaser, V. L., Pollock, N., & D'Adderio, L. (2021). The biography of an algorithm: Performing algorithmic technologies in organizations. *Organization Theory*, *2*(2), 26317877211004609.

Goh, Z.A.G. et al., 2021. Development of a novel method to investigate well-being and job quality in job tasks. In Singapore Conference of Applied Psychology 2021. Singapore Conference of Applied Psychology .

Goh, Z.A.G., Poon, K.W. (2022). A Task-Based Approach to Lifelong Learning, Well-Being, and Resilience in the Workplace of the Future. In: Lee, W.O., Brown, P., Goodwin, A.L., Green, A. (eds) International Handbook on Education Development in Asia-Pacific. Springer, Singapore. https://doi.org/10.1007/978-981-16-2327-1_72-1

Goos, M. and Manning, A., 2007. Lousy and lovely jobs: The rising polarization of work in Britain. *The review of economics and statistics*, *89*(1), pp.118-133.

Grønsund, T., & Aanestad, M. (2020). Augmenting the algorithm: Emerging human-in-the-loop work configurations. *The Journal of Strategic Information Systems*, *29*(2), 101614

Hadjimichael, D., & Tsoukas, H. (2019). Toward a better understanding of tacit knowledge in organizations: Taking stock and moving forward. *Academy of Management Annals*, *13*(2), 672-703.

Kellogg, K. C., Valentine, M. A., & Christin, A. (2020). Algorithms at work: The new contested terrain of control. *Academy of Management Annals*, *14*(1), 366-410.

Mayer-Schönberger, V., & Cukier, K. (2013). *Big data: A revolution that will transform how we live, work, and think*. Houghton Mifflin Harcourt.

Nott, M.T., Chapparo, C. and Heard, R., 2009. Reliability of the perceive, recall, plan and perform system of task analysis: A criterion-referenced assessment. *Australian Occupational Therapy Journal*, *56*(5), pp.307-314.

O*NET OnLine, National Center for O*NET Development (2022) Browse by Work Activities. Available at: www.onetonline.org/find/descriptor/browse/4.A. (Accessed: 29 April 2022).

Orlikowski, W.J., 2000. Using technology and constituting structures: A practice lens for studying technology in organizations. *Organization science*, *11*(4), pp.404-428.

Pachidi, S., Berends, H., Faraj, S., & Huysman, M. (2021). Make way for the algorithms: Symbolic actions and change in a regime of knowing. *Organization Science*, *32*(1), 18-41.

Pakarinen, P., & Huising, R. (2023). Relational Expertise: What Machines Can't Know. *Journal of Management Studies*.

Pentland, B.T., 2016. ThreadNet: Tracing and visualizing associations between actions. *IFIP Working Group 8.2, Dublin, Ireland*.

Peri, G. and Sparber, C., (2009). Task specialization, immigration, and wages. *American Economic Journal: Applied Economics*, *1*(3), pp.135-69.

Poon, K.W., Willems, T., Liu, W.S.Y. (2023). The Future of Expertise: From Stepwise Domain Upskilling to Multifaceted Mastery. In: Lee, W.O., Brown, P., Goodwin, A.L., Green, A. (eds) International

Handbook on Education Development in Asia-Pacific. Springer, Singapore. https://doi.org/10.1007/978-981-16-2327-1_42-1

Raisch, S., & Krakowski, S. (2021). Artificial intelligence and management: The automation–augmentation paradox. Academy of Management Review, 46(1), 192-210.

Schlogl, L., Weiss, E., & Prainsack, B. (2021). Constructing the 'Future of Work': An analysis of the policy discourse. *New Technology, Work and Employment*, *36*(3), 307-326.

Schwab, K. (2017). *The fourth industrial revolution*. Crown Currency.

Sergeeva, A. V., Faraj, S., & Huysman, M. (2020). Losing touch: an embodiment perspective on coordination in robotic surgery. *Organization Science*, *31*(5), 1248-1271.

Suchman, L. A. (2007). *Human-machine reconfigurations: Plans and situated actions*. Cambridge university press.

Susskind, D. (2020). *A world without work: Technology, automation and how we should respond*. Penguin UK.

Susskind, R. E., & Susskind, D. (2015). *The future of the professions: How technology will transform the work of human experts*. Oxford University Press, USA.

The National Center For O*NET Development (2003) Summary Report: Updating the Detailed Work Activities. Available at: https://www.onetcenter.org/reports/DWA_summary.html (Accessed: 29 April 2022).

Urry, J. (2016). What is the Future?. Cambridge: Polity.

Waardenburg, L., Huysman, M., & Sergeeva, A. V. (2022). In the land of the blind, the one-eyed man is king: Knowledge brokerage in the age of learning algorithms. *Organization Science*, *33*(1), 59-82

Wajcman, J. (2017). Automation: is it really different this time?. *The British Journal of Sociology*, *68*(1), pp. 119-127.

Willems, T., & Hafermalz, E. (2021). Distributed seeing: Algorithms and the reconfiguration of the workplace, a case of 'automated' trading. *Information and Organization*, *31*(4), 100376.